\documentclass[%
onecolumn,%
oneside,%
floats,%
aps,%
prd,%
nobibnotes,%
nofootinbib,%
amsmath,%
amssymb,%
amsfonts,%
amscd,%
superscriptaddress,%
eqsecnum%
]{revtex4}

\usepackage[utf8]{inputenc}
\usepackage{graphicx,array,dcolumn}
\usepackage[paperwidth=210mm,paperheight=297mm,centering,hmargin=1.8cm,vmargin=2.5cm]{geometry}
\newcommand{\bi}{\bibitem}
\newcommand{\be}{\begin{equation}}
\newcommand{\ee}{\end{equation}}

\def\b{\textbf}

\def\be{\begin{eqnarray}}
\def\ee{\end{eqnarray}}
\def\mpl{m_{Pl}}

\def\-g{\sqrt{-g}}

\renewcommand\rho{\varrho}
\renewcommand\tilde{\widetilde}

\begin{document}

\title{Spherically Symmetric Solutions in $F(R)$ Gravity and Gravitational Repulsion}

\author{E.V. Arbuzova}
\email{arbuzova@uni-dubna.ru}
\affiliation{Novosibirsk State University, Novosibirsk, 630090, Russia}
\affiliation{Department of Higher Mathematics, University "Dubna", 141980 Dubna, Russia}

\author{A.D. Dolgov}
\email{dolgov@fe.infn.it}
\affiliation{Novosibirsk State University, Novosibirsk, 630090, Russia}
\affiliation{ITEP, Bol. Cheremushkinsaya ul., 25, 113259 Moscow, Russia}
\affiliation{Dipartimento di Fisica e Scienze della Terra, Universit\`a degli Studi di Ferrara\\
Polo Scientifico e Tecnologico - Edificio C, Via Saragat 1, 44122 Ferrara, Italy}
\affiliation{Istituto Nazionale di Fisica Nucleare (INFN), Sezione di Ferrara\\
Polo Scientifico e Tecnologico - Edificio C, Via Saragat 1, 44122 Ferrara, Italy}

\author{L. Reverberi}
\email{reverberi@fe.infn.it}
\affiliation{Dipartimento di Fisica e Scienze della Terra, Universit\`a degli Studi di Ferrara\\
Polo Scientifico e Tecnologico - Edificio C, Via Saragat 1, 44122 Ferrara, Italy}
\affiliation{Istituto Nazionale di Fisica Nucleare (INFN), Sezione di Ferrara\\
Polo Scientifico e Tecnologico - Edificio C, Via Saragat 1, 44122 Ferrara, Italy}

\begin{abstract}
Spherically symmetric solutions in $F(R)$ theories in astronomical systems with rising energy density are 
studied. The range of parameters is established for which the flat space-time approximation for the 
background metric is valid. For the solutions in which the curvature scalar oscillates with large amplitude and high frequency, found in our previous papers, it is shown that the analysis of the Jeans instability is strongly modified. 
It is discovered that for large astronomical objects modified gravity can be repulsive, so such objects
shrink forming relatively thin shells instead of quasi uniform bodies. This may explain the formation of cosmic voids.
\end{abstract}

\maketitle

\section{Introduction}

The observed accelerated cosmological expansion stimulated an active study of gravity modifications
at large scales which could lead to such anti-gravitation in the absence of matter. To this end a non-linear function
of the curvature scalar, $F(R)$, is added to the the standard Einstein-Hilbert action of General
Relativity (GR):
\be
S = \frac{m_{Pl}^2}{16\pi} \int d^4 x \sqrt{-g}\,[R +F(R)]+S_m, 
\label{A1}
\ee 
where $m_{Pl}= 1.22\cdot 10^{19}$ GeV is the Planck mass, 
$R$ is the scalar curvature, and $S_m$ is the matter action. This approach was pioneered in works~\cite{grav-mdf}
and the flux of papers on the subjects looks unabated. For a review of these theories see e.g.~\cite{F-of-R-rev}.

The corresponding equations of motion have the form:
\be
\left( 1 + F_{,R}\right) R_{\mu\nu} -\frac{1}{2}\left( R + F\right)g_{\mu\nu}
+ \left( g_{\mu\nu} D_\alpha D^\alpha - D_\mu D_\nu \right) F_{,R}  = 
\frac{8\pi T_{\mu\nu}}{m_{Pl}^2}\,,
\label{eq-of-mot}
\ee 
where $F_{,R} = dF/dR$, $D_\mu$ is the covariant derivative, and 
$T_{\mu\nu}$ is the energy-momentum tensor of matter.
Taking the trace of these equations we obtain the following equation for the curvature scalar:
\be
3 D^2 F_{,R} -R + R F_{,R} - 2F = \tilde T\,,
\label{D2-R}
\ee
where $\tilde T = \tilde T_\mu^\mu $ and $ \tilde T_{\mu\nu} = 8\pi T_{\mu\nu}/m_{Pl}^2$.
Note that in the absence of matter this equation contains only $R$ and in this sense is closed.
However, $T_{\mu\nu}$ depends upon the metric and in the general case the total set of equations~(\ref{eq-of-mot})
is to be used. Note that in the usual GR, i.e. for $F(R) = 0$, curvature is expressed algebraically through the trace
of the energy-momentum tensor, $R = -\tilde T$, while for $F(R) \neq 0$ (even very small), deviations from
GR may be significant. This is the common case for differential equations with a small coefficient multiplying the highest derivative.

The function $F(R)$ is chosen in such a way that this equation, in the absence of matter, has solution $R=$ const.,  
corresponding to an accelerated universe expansion.

Popular $F(R)$ models phenomenologically acceptable for cosmology have been suggested in~\cite{Starob,HuSaw,ApplBatt}. They are more or less equivalent, particularly the former two, and in what follows we will use the specific $F(R)$ of ref.~\cite{Starob}:
\be
F(R) =  -\lambda R_0\left[1-\left(1+\frac{R^2}{R_0^2}\right)^{-n}\right] - \frac{R^2}{6m^2},
\label{F-of-R}
\ee
where $R_0$ is a constant parameter with dimensions of curvature and similar in magnitude to the cosmological curvature at the present day universe, $\lambda$ is a dimensionless constant of order unity and the power $n$ is usually taken to be an integer (though not necessarily so).
The last term is introduced to avoid infinite $R$ singularities in the past cosmology~\cite{App-Bat-Star}
or in the future in astronomical systems with rising energy density~\cite{future-sing}.  

A detailed study of the solutions of the modified gravity equations in the present day universe
was performed in ref.~\cite{ADR-1,ADR-2} for finite-size astronomical objects. 
It was found that if the energy density rises with time, fast oscillations of the scalar curvature are induced, with an amplitude possibly much larger than the usual GR value $R=-\tilde T$. The solution has the form:
\be
R = R_{GR} (r) y (t),
\label{R-of-t}
\ee
where $R_{GR} = - \tilde T(r)$ is the would-be solution in the limit of GR, while the quickly oscillating function
$y(t)$ may be much larger than unity. According to ref.~\cite{ADR-2} the maximum value of $y$ in the so-called spike region is:
\be
y(t) \sim 6 n (2n+1)  m t_u \left(\frac{t_u}{t_{contr} }\right) \left[\frac{\rho_m (t)}{\rho_{m0}} \right]^{(n+1)/2}
\left(\frac{\rho_c}{\rho_{m0}} \right)^{2n+2} ,
\label{y-of-t}
\ee 
where $t_u$ is the universe age, $t_{contr} $ is the characteristic contraction time, so the energy density of the contracting cloud behaves as $\rho_m (t) = \rho_{m0} (1 + t/t_{contr})$, with $\rho_{m0} $ being the initial energy density of the cloud, and $\rho_c~=~10^{-29}$~g/cm$^3$ being the present day cosmological energy density. According to ref.~\cite{ADR-cosm}, the mass parameter $m$ entering eq.~(\ref{F-of-R}) should be larger than about $10^5$ GeV to avoid a conflict with BBN. So the factor $m t_u $ is huge: $ m t_u \geq 10^{47}$.

As shown in ref.~\cite{ADR-2}, such high amplitude spikes are formed if 
\be
6 n^2 (2n+1)^2 \left(\frac{t_u}{t_{contr}}\right)^2 \left[\frac{\rho_m (t)}{\rho_{m0}} \right]^{3n+1}
\left(\frac{\rho_c}{\rho_{m0}} \right)^{2n+2} > 1.
\label{spike-region}
\ee
This condition is satisfied e.g. for $\rho_{m0}/\rho_c = 1-10$, $\rho_m(t)/\rho_{m0} =10-100$ and sufficiently large $n$.

\section{Spherically Symmetric Solutions in $F(R)$ Gravity}

The analysis of ref.~\cite{ADR-1,ADR-2} has been done under the assumption that the background 
space-time is nearly flat and so the background metric is almost Minkowsky. However, the large deviation of curvature from its GR value, found in these works, may invalidate the assumption of an approximately flat background and should be verified.\\
In what follows we consider a spherically symmetric bubble of matter, e.g. a gas cloud or some other astronomical object, which occupies a finite region of space of radius $r_m$, and study spherically symmetric solution of eqs.~(\ref{eq-of-mot}), assuming that the metric has the Schwarzschild form:
\be
ds^2 =  A (r,t) dt^2 - B(r,t) dr^2 - r^2 (d\theta^2 + \sin^2 \theta\,d\phi^2) .
\label{ds2}
\ee
 Metric of this type in $F(R)$ theories was analyzed in ref.~\cite{sch-dombriz} but the curvature oscillations, which are in the essence of
our work (see below), were not taken into account there. 

We assume that the metric coefficients $A$ and $B$ weakly deviate from unity and check when this is true. The nonzero components of the Ricci tensor corresponding to the metric~(\ref{ds2}) are:
\be
R_{00}&=&\frac{A''-\ddot B}{2B}+\frac{(\dot B)^2-A'B'}{4B^2}+\frac{\dot A \dot B - (A')^2}{4AB}+\frac{A'}{rB}\, , 
\label{R00}\\
R_{rr}&=&\frac{\ddot B -A''}{2A}+\frac{(A')^2-\dot A \dot B}{4A^2}+\frac{ A' B' - (\dot B)^2}{4AB}+\frac{B'}{rB}\, , 
\label{R11}\\
R_{\theta \theta }&=&-\frac{1}{B}+\frac{rB'}{2B^2}-\frac{rA'}{2AB}+1\, , 
\label{R22} \\
R_{\varphi \varphi }&=&\left(-\frac{1}{B}+\frac{rB'}{2B^2}-\frac{rA'}{2AB}+1\right )\sin^2\theta  = R_{\theta \theta} \sin^2\theta \, ,
\label{R33} \\
R_{0r}&=&\frac{\dot B}{rB}\, .
\label{R01}
\ee
Here a prime and an overdot denote differentiation with respect to $r$ and $t$, respectively. The corresponding Ricci scalar is equal to:
\be 
R&=&\frac{1}{A}R_{00}-\frac{1}{B}R_{rr}-\frac{1}{r^2}R_{\theta \theta}-\frac{1}{r^2\sin^2\theta }R_{\varphi \varphi}  \nonumber \\
&=&\frac{A''-\ddot B}{AB}+\frac{(\dot B)^2-A'B'}{2AB^2}+\frac{\dot A \dot B - (A')^2}{2A^2B}+\frac{2A'}{rAB}-\frac{2B'}{rB^2}+\frac{2}{r^2B}-\frac{2}{r^2}
\label{Rscalar}\\
&=& \frac{2}{A}R_{00}-\frac{2B'}{rB^2}+\frac{2}{r^2B}-\frac{2}{r^2} \, .
\nonumber 
\ee
We assume that the metric is close to the flat one, i.e.
\be
A_1 = A - 1 \ll 1\,\,\,{\rm and} \,\,\, B_1 = B-1 \ll 1
\label{A1-B1}
\ee
and study if and when this assumption remains true for the solutions with very large values of $R$ found in our previous works~\cite{ADR-1,ADR-2}. It is convenient to use equations (\ref{eq-of-mot}) in the following form:
\be 
R_{00}  - R/2 &=& \frac{ \tilde T_{00} + \Delta F_{,R} + F/2 - RF_{,R}/2}{1+F_{,R}},
\label{R-00} \\
R_{rr}  + R/2 &=&  \frac{ \tilde T_{rr} + (\partial_t^2 +\partial_r^2 - \Delta) F_{,R} - F/2 + RF_{,R}/2}{1+F_{,R}} ,
\label{R-rr}
\ee
because their  left hand sides contain only first derivatives of the metric coefficients. In the weak field limit, when derivatives of $A(r,t)$ and $B(r,t)$ are sufficiently small so that their square can be neglected, we obtain the following expressions for the $R_{00}$ and $R_{rr}$ components of the Ricci tensor and for the Ricci scalar $R$:
\be
R_{00}&\approx&\frac{A''-\ddot B}{2}+\frac{A'}{r}\, , 
\label{R00-weak}\\
R_{rr}&\approx&\frac{\ddot B -A''}{2}+\frac{B'}{r}\, , 
\label{R11-weak}\\
R&\approx&A''-\ddot B+\frac{2A'}{r}-\frac{2B'}{r}+\frac{2(1-B)}{r^2}\, .
\label{R-weak}
\ee
If the energy density of matter inside the the cloud, i.e. for  $ r<r_m$, is much larger than the cosmological energy density,
the following restrictions are fulfilled: 
\be
F_{,R} \ll 1\,\,\,\,\, {\rm and}\,\,\,\,\,\, F \ll R
\label{bounds-on-F}
\ee 
For static solutions the effects of gravity modifications in this limit are weak and, as we will see in what follows, the solution is quite close to the standard Schwarzschild one. We also assume for simplicity that the 
spatial derivatives of $F_{,R}$ are small in comparison with the time derivatives, so  from eq.~(\ref{D2-R}) it follows that $(\partial_t^2 - \Delta ) F_{,R} = (\tilde T + R)/3 $ and we find:
\be
B_1' + \frac{B_1}{r} & = & r \tilde T_{00},
\label{B1-prime} \\
A''_1-\frac{A_1'}{r} &  = & -\frac{3B_1}{r^2} +\ddot B_1 +\tilde T_{00}-2\tilde T_{rr} + \frac{\tilde T_{\theta\theta}}{r^2} +
 \frac{\tilde T_{\varphi\varphi}}{r^2\sin^2\theta}
\equiv S_A\, .
\label{A1-two-prime}
\ee
Since we assumed small deviations from the Minkowsky metric, we neglected the corresponding corrections in $T_{\mu \nu}$. The validity of this assumption is precisely what we have to check.

Equation~(\ref{B1-prime}) has the solution:
\be 
B_1(r,t) = \frac{C_B(t)}{r}+\frac{1}{r}\,\int_0^r dr' r'^2 \tilde T_{00} (r',t)\, .
\label{B1-new}
\ee
To avoid a singularity at $r=0$ we have to assume that $C_{B}(t) \equiv 0$. Then this expression for $B_1$ formally coincides with the usual Schwarzschild solution, while the equation determining the metric coefficient $A_1$ allows for an additional freedom:
\be 
A_1(r,t)=C_{1A}(t)r^2+C_{2A}(t)+\int^{r_m}_r dr_1\,r_1 \int^{r_m}_{r_1} \frac{dr_2}{r_2}\, S_A(r_2,t)\, .
\label{A1-common}  
\ee
The integration limits are chosen in such a way that the singularity at $r_2=0$ is avoided. Using equation (\ref{B1-new}) with $C_{B}=0$ we can rewrite $S_A$ as:
\be 
S_A = -\frac{3}{r^3}\,\int_0^r dr' r'^2 \tilde T_{00} (r',t) 
+\frac{1}{r}\,\int_0^r dr' r'^2 \ddot {\tilde {T}}_{00} (r',t) \, + 
\tilde T_{00}-2\tilde T_{rr} + \frac{\tilde T_{\theta\theta}}{r^2} +
 \frac{\tilde T_{\varphi\varphi}}{r^2\sin^2\theta}\, .
\label{S-A}
\ee
Accordingly we obtain the following expression for $A_1(r,t)$:
\be \nonumber
 A_1(r,t) = C_{1A}(t)r^2+C_{2A}(t)
+\int_{r}^{r_m} dr_1\,r_1\int_{r_1}^{r_m} \frac{dr_2}{r_2}\, \left(\tilde T_{00} (r_2,t)-2\tilde T_{rr}(r_2,t) +  \frac{\tilde T_{\theta\theta}(r_2,t)}{r^2} +
\right.\\  %\nonumber
 \left.  \frac{\tilde T_{\varphi\varphi}(r_2,t)}{r^2\sin^2\theta}\right)  % \\ %\nonumber
-\int_{r}^{r_m} dr_1\,r_1\int_{r_1}^{r_m} \frac{dr_2}{r_2} \, \left(\frac{3}{r_2^3}\,\int_0^{r_2} dr' r'^2 \tilde T_{00} (r',t) -
\frac{1}{r_2}\,\int_0^{r_2} dr' r'^2 \ddot {\tilde {T}}_{00} (r',t)\right).
\label{A-1}
\ee

\subsection{The Schwarzschild Limit}

It is instructive to check how solutions (\ref{B1-new}) and (\ref{A-1}) reduce to the vacuum Schwarzschild solution in GR. The mass of matter inside a radius $r$ is defined in the usual way:
\be
M (r,t) =\int_0^{r} d^3r\,  T_{00} (r,t)=4\pi \int_0^{r} dr\,  r^2\,  T_{00} (r,t)
\label{M}
\ee
If all matter is confined inside a radius $r_m$, the total mass is $M \equiv M(r_m)$ and due to mass conservation it does not depend on time. Since $\tilde T_{00}=8\pi T_{00}/m_{Pl}^2$, we obtain for $r>r_m$, as expected, $B_1=r_g/r$, where $r_g=2M/m_{Pl}^2$ is the usual Schwarzschild radius. 
 
Let us turn now to the calculation of $A_1$ (\ref{A-1}). Evidently, for $r>r_m$ the first integral term vanishes because $r_2$ is also larger than $r_m$, in fact in this region we have $T_{\mu \nu}=0$. The integral containing $\ddot {\tilde {T}}_{00} $ is also zero due to total mass conservation. The remaining integral can be easily taken:
\be
\int_{r}^{r_m} dr_1\,r_1\int_{r_1}^{r_m} \frac{dr_2}{r_2} \, \frac{3}{r_2^3}\,\int_0^{r_2} dr' r'^2 \tilde T_{00} (r',t) = 
\frac{r_g}{r}+\frac{r_g\,r^2}{2r_m^3}-\frac{3r_g}{2r_m}\, .
\label{int-A}
\ee
Thus the metric coefficient outside the source is:
\be 
A_1 = -\frac{r_g}{r}+\left[C_{1A}(t)-\frac{r_g}{2r_m^3}\right] r^2 +\left[C_{2A}(t)+\frac{3r_g}{2r_m}\right] \, .     
\label{A-1-Sch}
\ee
Choosing $C_{1A}=r_g/(2r_m^3)$, to eliminate the $r^2$-term at infinity, and $C_{2A}=-3r_g/(2r_m)$ we obtain the usual Schwarzschild solution. Note that it is not necessary to demand that the space-independent constant in $A_1$ must vanish, because it can be removed by a redefinition of the time variable.

\subsection{Modified Gravity Solutions}

In the modified theory the internal solution remains of the same form (\ref{B1-new}) and (\ref{A-1}), where the coefficient $C_{1A}$, however, may depend non-trivially on time. This coefficient can be found from eq. (\ref{R-weak}) if the curvature scalar is known. As previously mentioned, we have shown in papers \cite{ADR-1, ADR-2} in systems with rising energy density that the curvature scalar may be much larger than its value in GR. Using eqs. (\ref{B1-new}) and (\ref{A-1}) and comparing them to eq. (\ref{R-weak}) we can conclude that the dominant contribution into such form of the curvature is given by $A''+2A'/r$, i.e.  $C_{1A}(t)=R(t)/6$, where $R(t)$ is given by eqs. (\ref{R-of-t},\ref{y-of-t}).

There is an essential difference between the modified and the standard solutions in vacuum. 
In the standard case the term proportional to $r^2$ appears both at $r<r_m$ and $r>r_m$ with the same coefficient and 
hence it must vanish. On the other hand, for modified gravity such condition is not applicable and the $C_{1A} r^2$-term may be present at $r<r_m$ and absent at $r \gg r_m$. The vacuum solution for $R$ is presumably $R\sim R_c$, where $R_c$ the small cosmological curvature, plus possible oscillating terms.

Thus to summarize, the metric functions inside the cloud are equal to:
\be
B (r, t) & =& 1 + \frac{2M(r,t)}{\mpl^2r} \equiv  1+ B_1^{(Sch)}\, , \label{B-of-r-t} \\ 
A(r,t)  &=& 1 + \frac{R(t )\,r^2}{6} + A_1^{(Sch)} (r,t) \label{A-of-r-t}\, .
\ee
In other words we construct the internal solution assuming that it consists of two terms: the Schwarzschild one and the oscillating part generated by the rising density as is shown in our works~\cite{ADR-1, ADR-2}. The expression for $A_1^{(Sch)} (r,t) $ can be found from (\ref{A-1}) with constant $C_{A1}=r_g/2r_m^3$ and $C_{A2}=-3r_g/r_m$, as determined from eq.~(\ref{A-1-Sch}). As for the integrals in eq.~(\ref{A-1}), we calculated them assuming that matter is nonrelativistic, so the space components of $T_{\mu\nu}$ are negligible in comparison to $T_{00}$, and that the matter/energy density, $T_{00} \equiv \rho_m (t)$, is spatially constant but may depend on time. 
The first two integrals in eq. (\ref{A-1}) cancel out and only the integral containing the second time derivative of the mass density survives. So for the Schwarzschild part of the solution we find: 
\be
A_1^{(Sch)} (r,t) = \frac{r_g r^2}{2r_m^3} -\frac{3r_g}{2r_m}+
 \frac{ \pi \ddot\rho_m}{3 m_{Pl}^2}\, ( r_m^2 - r^2)^2\, .
\label{A-1-2}
\ee

Since $R(t)$ is much larger that its GR value, $|R_{GR}|=8 \pi \rho_m /m_{Pl}^2$, the oscillating part $R(t)r^2/6$ gives the dominant contribution into $A_1$. Indeed, $r^2 R(t) \sim r^2 y R_{GR}$ with $y \gg 1$, while the canonical Schwarzschild terms are of the order of $r_g/r_m \sim \rho_m r_m^2/m_{Pl}^2 \sim r_m^2 R_{GR} $. If the initial energy density of the cloud is of the order of the cosmological energy density, i.e. $R_{GR}\sim 1/t_u^2$, then the metric would noticeably deviate from the Minkowsky metric for clouds having radius $r_m > t_u/\sqrt y$, where $y$ is given by eq. (\ref{y-of-t}). This is a realistic situation because structure formation proceeds at red shifts of order unity when the density fluctuations $\delta \rho $ became of the same order of the background cosmological energy density. 

For large objects, such that $R r^2 /6 \sim 1$, the approximation used here is not applicable and one has to solve the exact non-linear equations with $R_{\mu\nu}$ given by eqs.~(\ref{R00}-\ref{R01}); this situation will be studied elsewhere. If $A_1$ becomes comparable with unity, the evolution of $R(t)$ may significantly differ from that 
found in~\cite{ADR-1,ADR-2}, but even for small $A_1$ there would arise interesting new effects considered
below. 

In the lowest order in the gravitational interaction the motion of a non-relativistic  test particle is governed by the equation:
\be
\ddot r = - \frac{A'}{2} = -\frac{1}{2}\left[ \frac{R(t) r}{3} + \frac{r_g r}{r_m^3} \right],
\label{ddot-r}
\ee
where $A$ is given by eq. (\ref{A-of-r-t}). Since $R(t)$ is always negative and large, the modifications of GR considered here lead to anti-gravity inside a cloud with energy density exceeding the cosmological one. Gravitational repulsion dominates over the usual attraction if
\be
\frac{|R|r_m^3}{3r_g} = \frac{|R|r_m^3m_{Pl}^2}{6 M} = \frac{|R|r_m^3m_{Pl}^2}{8\pi \rho\,r_m^3} = \frac{|R|}{\tilde T_{00}} > 1\,,
\ee
so basically whenever $R$ starts oscillating, regardless of the initial value of $\rho$ and to some extent of the specific $F(R)$ considered. Therefore, this is most likely a more fundamental statement, applicable to essentially all $F(R)$ models producing oscillations of $R$.

\section{Conclusions}

As it was shown in ref.~\cite{ADR-1,ADR-2}, the time evolution of curvature exhibits high narrow spikes over some smooth background with relatively low $R$ -- see e.g. eq.~(4.17) of~\cite{ADR-2}. These spikes are damped due to gravitational particle production but the corresponding life-time could be comparable or even larger than the cosmological time. So structure formation in modified gravity would be very much different from that in the standard GR. Sufficiently large primordial clouds would not shrink down to smaller and smaller bodies with more or less uniform density but form thin shells empty (or almost empty) inside. { This result is in clear contrast with recent studies of the formation and stability of astronomical structures in $F(R)$ gravity~\cite{struct-formation}; we should however stress that those works neglected time derivatives, so the contrast with our results is mainly due to the different physical phenomena involved.} This anti-gravitating behavior may also be a possible driving force for the creation of cosmic voids.

\acknowledgements
EA and AD acknowledge the support of the grant of the Russian Federation government 11.G34.31.0047. \\[4mm]


\begin{thebibliography}{99}

\bibitem{grav-mdf} 
S. Capozziello, S. Carloni, A. Troisi,
Recent Res. Dev. Astron. Astrophys. {\bf 1}, 625 (2003);
arXiv:astro-ph/0303041;\\
S.M. Carroll, V. Duvvuri, M. Trodden, M.S. Turner,
Phys.Rev. D {\bf 70}, 043528 (2004); arXiv:astro-ph/0306438.

\bi{F-of-R-rev}
T. Clifton, P.G. Ferreira, A. Padilla, C. Skordis, Physics Reports {\bf 513}, 1 (2012), 1-189 [arXiv:1106.2476 [astro-ph.CO]];\\
S.~'i.~Nojiri and S.~D.~Odintsov,
  %``Unified cosmic history in modified gravity: from F(R) theory to Lorentz non-invariant models,''
  Phys.\ Rept.\  {\bf 505} (2011) 59
  [arXiv:1011.0544 [gr-qc]].


\bibitem{Starob} 
A.A. Starobinsky, JETP Lett. {\bf 86}, 157 (2007).

\bibitem{HuSaw} 
W. Hu, I.  Sawicki, Phys. Rev. D {\bf 76}, 064004 (2007).

\bibitem{ApplBatt} 
A. Appleby, R.  Battye, Phys. Lett. B {\bf 654}, 7 (2007).

\bi{App-Bat-Star}
S.A. Appleby, R.A. Battye, A.A. Starobinsky, {\it JCAP} \b{1006} (2010) 005.

\bi{future-sing}
A.~V.~Frolov,
  %``A Singularity Problem with f(R) Dark Energy,''
  Phys.\ Rev.\ Lett.\  {\bf 101} (2008) 061103
  [arXiv:0803.2500 [astro-ph]];\\
E.~V.~Arbuzova and A.~D.~Dolgov,
  %``Explosive phenomena in modified gravity,''
Phys.\ Lett.\ B {\bf 700} (2011) 289 [arXiv:1012.1963 [astro-ph.CO]];\\
L. Reverberi, Phys. Rev. D {\bf 87}, 084005 (2013) [arXiv:1212.2870 [gr-qc]].

\bi{ADR-1} 
E.V. Arbuzova, A.D. Dolgov, L. Reverberi, {Eur. Phys. J. C} {\bf 72} (2012) 2247.

\bi{ADR-2}
E.V. Arbuzova, A.D. Dolgov, L. Reverberi, arXiv:1305.5668.

\bi{sch-dombriz}
A. de la Cruz-Dombriz, A. Dobado, A. L. Maroto, Phys. Rev. {\bf D80}, 124011 (2009).

\bi{ADR-cosm}
E.V. Arbuzova, A.D. Dolgov, L. Reverberi, {JCAP} \b{02} (2012) 049.

\bibitem{struct-formation}
  S.~Capozziello, M.~De Laurentis, S.~D.~Odintsov and A.~Stabile,
  %``Hydrostatic equilibrium and stellar structure in f(R)-gravity,''
  Phys.\ Rev.\ D {\bf 83} (2011) 064004
  [arXiv:1101.0219 [gr-qc]];\\
 S.~Capozziello, M.~De Laurentis, I.~De Martino, M.~Formisano and S.~D.~Odintsov,
  %``Jeans analysis of self-gravitating systems in f(R)-gravity,''
  Phys.\ Rev.\ D {\bf 85} (2012) 044022
  [arXiv:1112.0761 [gr-qc]].


\end{thebibliography}
\end{document}